# Local Dielectric Measurements of BaTiO$_3$-CoFe$_2$O$_4$ Nano-composites Through Microwave Microscopy

Yi Qi[1)], H. Zheng[2)], R. Ramesh[2)], Steven M. Anlage[1), *]

1) Materials Research Science and Engineering Center, Departments of Physics and Materials Science and Engineering,

University of Maryland, College Park

2) Department of Materials Science and Engineering, University of California at Berkeley

* Address all correspondence to Steven M. Anlage, e-mail: anlage@squid.umd.edu or Yi Qi, email: yiqi@umd.edu

We report on linear and non-linear dielectric property measurements of BaTiO$_3$ – CoFe$_2$O$_4$ (BTO-CFO) ferroelectro-magnetic nano-composites and pure BaTiO$_3$ and CoFe$_2$O$_4$ samples with Scanning Near Field Microwave Microscopy. The permittivity scanning image with spatial resolution on the μm scale shows that the nano-composites have very uniform quality with an effective dielectric constant $\varepsilon_r$ = 140 ± 6.4 at 3.8 GHz and room temperature. The temperature dependence of dielectric permittivity shows that the Curie temperature of pure BTO was shifted by the clamping effect of the MgO substrate, whereas the Curie temperature shift of the BTO ferroelectric phase in BTO-CFO composites is less pronounced, and if it exists at all, would be mainly caused by the CFO. Non-linear dielectric measurements of BTO-CFO show good ferroelectric properties from the BTO.

**Introduction**

Ferroelectro-magnets are materials in which magnetic and electric dipolar orderings coexist. Because of the additional degree of freedom, there is current interest in using the materials for applications such as memory, electric field-controlled ferromagnetic resonance devices, magnetically switched electro-optical devices, etc.[1]. Besides single phase ferroelectro-magnetic materials[2-3], researchers have developed multiphase materials consisting of layers or mixtures of piezoelectric and magnetostrictive phases[4-5].

In the nano-pillar-in-matrix kind of multiphase materials introduced by H. Zheng et al.[6], the properties strongly depend on the pillar size and composition, as well as the uniformity. A localized dielectric measurement is a convenient and effective method to check the quality of the film and to find the nanostructure/composition dependence of the electrical properties. In this paper, we introduce localized dielectric measurements of BTO-CFO using a microwave microscope. We report the measurement of relative dielectric constant with good sensitivity (about ±4 in the range of 200~1000) and with high spatial resolution (~μm). Another motivation of this paper is to investigate the change in ferroelectric properties of the BTO due to the introduction of the secondary phase CFO. We measured the temperature dependence and electric field tunability of the dielectric properties on these new materials and compare with a pure BTO thin film. The results show evidence that the BTO dielectric properties are significantly affected by the CFO pillars.

**Experiments**

Our scanning microwave microscope consists of an open-ended coaxial probe with a sharp, protruding center conductor[7]. The sample under the tip perturbs the resonator and causes a frequency shift. The frequency shift from the unperturbed condition (with a known sample as reference) to the perturbed condition (with the unknown sample to be measured) is related to the permittivity of the sample. Using perturbation theory[8], we calculated the frequency shift of the microscope as a function of the fields in the sample near the probe tip,[7]

$$\frac{\Delta f}{f} \approx \frac{\varepsilon_0}{4W} \int_{V_s} (\varepsilon_{r2} - \varepsilon_{r1}) \vec{E}_1 \bullet \vec{E}_2 dV ,$$

where $\varepsilon_{r2}$ and $\varepsilon_{r1}$ are the relative permittivity of the two samples. $E_1$ and $E_2$ are the calculated electric fields inside the two samples, W is the energy stored in the resonator, and the integral is over the volume $V_s$ of the sample.

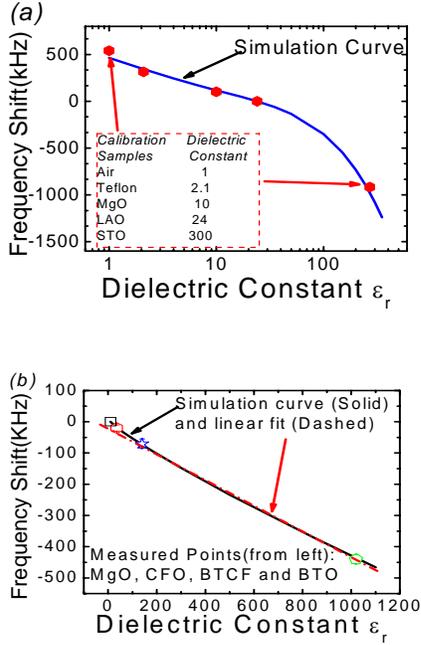

Fig 1. (a) Data and simulation curve of frequency shift vs. permittivity for bulk samples (we use a bulk LAO substrate as the reference and all frequency shifts are given relative to it). The plot is in a liner-log scale to show the diversity of dielectric reference points. The relationship is non-linear for relative dielectric constant values below 50, but approximately linear above 50; (b) data and simulation for thin films. We assume MgO as the substrate and a film thickness of 180 nm. All frequency shifts are given relative to MgO. Note that this plot is in a linear scale.

The dielectric constant vs. frequency shift curve for bulk and thin film samples are illustrated in Fig 1. The shape of curve is determined by the probe geometry. Here we simplify the probe geometry as an ellipsoid of revolution with a blunt end (illustrated in Fig2. (c)). The geometry can be modified by varying three parameters, i.e. long axis $a$, short axis $b$ and blunt end width $c$. For a defined geometry, we use Maxwell 2D software (Ansoft) to calculate the static electric field and stored energy in the sample. It is established that the static fields are a good approximation to the near-field structure of the tip.[7] Combined with the equation above, the frequency shift vs. permittivity curve is calculated. To find the probe geometry of the tip used in our experiment, i.e. the correct $a$, $b$, and $c$ parameters, calibration points were measured with several known bulk samples. Fig 1.(a) shows a calculated curve of frequency shift vs. permittivity that fits our calibration points well when we choose a=50μm, b=22μm and c=2.6μm. The geometry parameters are confirmed by SEM observation of the tip [9].

We can see that larger dielectric permittivity causes a more negative frequency shift of the microscope.

Fig 1.(b) simulates the thin film situation with the probe geometry determined by the bulk dielectric calibration step. In this simulation, we assume the substrate is MgO and the thin film thickness is 180nm, which is consistent with the samples used in this experiment. We assumed the sample has a disk shape with a diameter of 5000μm and a substrate thickness of 500μm. In Fig 1.b we define the bare MgO substrate as the unperturbed condition, the frequency shift relative to MgO is recorded to get the dielectric permittivity for the thin film samples to be measured. The simulation curve is close to a straight line in the range from $\varepsilon_r$=100 to $\varepsilon_r$=1100 ($\varepsilon_r \approx$ -15.425 - 2.336 $\Delta f$ (kHz)), so we can use the linear fit to estimate the dielectric constant of the thin films.

**Samples**

The BTO-CFO thin film nano-composite material is prepared by pulsed-laser deposition on MgO substrates. All films discussed here have a nominal thickness of 180nm. X-ray diffraction and TEM experiments show that the BTO and CFO were separated into two phases in the thin film by self-assembly [6]. The two phases have a pillar-in-matrix geometry, in which the CFO pillars occupy about 35% in volume and have diameters of 20-30 nm, and their distribution and size are roughly uniform. Both BTO and CFO phases have their [100] lattice direction perpendicular to the thin film plane [6].

Two types of BTO-CFO samples were prepared (see Table I). One without a counter-electrode for linear dielectric measurement (a) and another one with a counter-electrode $SrRuO_3$ (SRO) layer for non-linear dielectric measurement (b). The sheet resistance of the SRO layer is 2700 Ω/square, sufficiently large to render it effectively invisible to the microwave signal to good approximation.[10] For the former samples, part of the thin film was ion milled to expose the MgO substrate (see Fig 2.a). Again, we use the MgO substrate as the unperturbed condition and the BTO-CFO on MgO as the perturbed condition, while measuring the frequency shift. For the latter case, we use a bias tee outside the inductively de-coupled resonator to apply bias voltage between the SRO counter-electrode and the microscope tip to get the dielectric response vs. bias voltage relationship (Fig 2.b) [7]. For comparison purposes, pure BTO and pure CFO thin film samples were also prepared with the same geometry and lattice direction, that is BTO on MgO (c) and BTO on SRO counterelectrode on MgO substrate(d), CFO on MgO (e) and CFO on SRO counterelectrode on MgO substrate(f). The samples prepared for this study are listed in Table I.

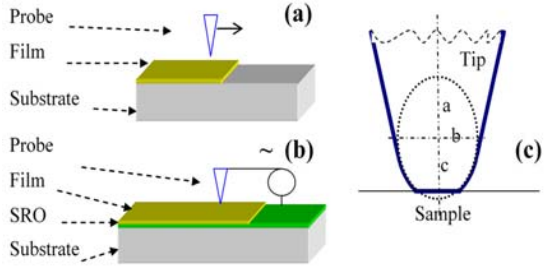

Fig 2. (a) Sample schematic for dielectric measurements (BTO-CFO, pure BTO or pure CFO thin film directly on MgO substrate with partly exposed MgO) (b) Sample schematic for non-linear dielectric measurements (BTO-CFO, pure BTO or pure CFO thin film on MgO substrate with a SRO counter-electrode layer). A dc and low frequency (Hz~kHz) ac voltage can be applied across the dielectric film between the tip and the counterelectrode. (c) Simplified geometry of the probe tip modeled as an ellipsoid of revolution in contact with the sample.

| Sample Index | a | b | c | d | e | f |
|---|---|---|---|---|---|---|
| Material | BTO-CFO | BTO-CFO | BTO | BTO | CFO | CFO |
| SRO Counter-electrode | NO | YES | NO | YES | NO | YES |
| $\varepsilon_{avg}$ from linear measurements (room temperature, 3.8 GHz) | 140 ± 6, | | 1020 ± 20 | | 35 ± 2 | |
| $T_{max}$ from linear measurements | 124 °C | | 154 °C | | N/A | |
| $T_c$ from non-linear measurements | | 130 °C | | 130 °C | | N/A |

Table I. Summary of samples measured and their dielectric properties. All films are grown on MgO substrates.

**Results and Discussions**
**1. Quantitative Dielectric Imaging Resolution**
We scanned the microwave microscope tip across the lithographically-defined sharp edge between the BTO-CFO thin film and MgO bare substrate and recorded the frequency shift. We use this design rather than measuring two separate samples because it can reduce the systematic error and noise in the dielectric constant determination, especially for the temperature dependence measurements presented below. There is a sharp change in frequency shift upon crossing the edge from MgO to BTO-CFO or BTO (Fig 3). The negative frequency change from MgO to BTO-CFO or BTO indicates BTO-CFO and BTO have higher permittivity than MgO ($\varepsilon_r$ ~10). We can convert the frequency shift values to relative permittivity values of $\varepsilon_r$ = 140 ± 6, 1020 ± 20 and 35 ± 2 for BTO-CFO, BTO and CFO respectively (at room temperature and 3.8 GHz). Assuming that the patterned edge and dielectric properties have a sharp change at the edge, from the scan with 0.1 μm step size (inset of Fig 3. (b)), we can also see the microwave microscope has a quantitative resolution[11] for dielectric constant of about 5 μm.

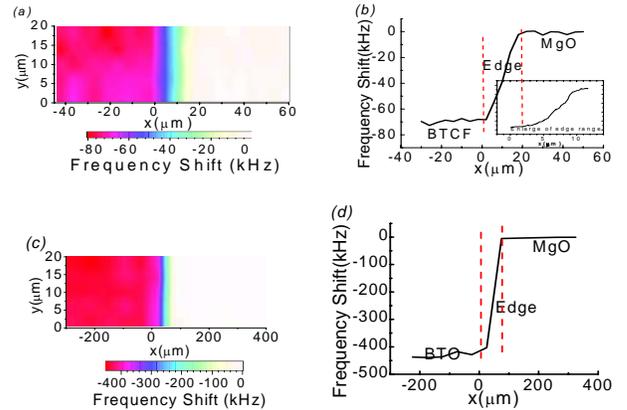

Fig 3. (a), (c) Frequency shift scanning images at 3.8 GHz and room temperature crossing the BTO-CFO/MgO edge and BTO/MgO edge respectively, (b) (d) Single line scans from (a) and (c), inset of (b) is a scan across the BTO-CFO/MgO edge with 0.1 μm step size.

**2. Linear Dielectric Imaging**
A two-dimensional scanning image [Fig 4. (a)] shows that the BTO-CFO has a fairly uniform dielectric property with $\varepsilon_r$ = 140 ± 6.4 averaged over a 2mm x 2mm area. The dielectric properties of this kind of nano-composite mainly depend on composition [5]. From TEM and AFM measurement of this sample [6], we see it has roughly uniform composition in the plane of the film. That may explain the small range of dielectric constant variation observed. Fig 4. b shows a finer scanning image of the most inhomogeneous part of the film with 0.5 μm step size. The dielectric constant is $\varepsilon_r$ = 136 ± 8.2 over this area. There is still contrast in the dielectric image on this short length scale, indicating a "qualitative" spatial resolution [11] on the μm scale. However, the resolution for quantitative imaging is larger and is typically governed by the tip geometry[9] and the field confinement volume in the sample [11].

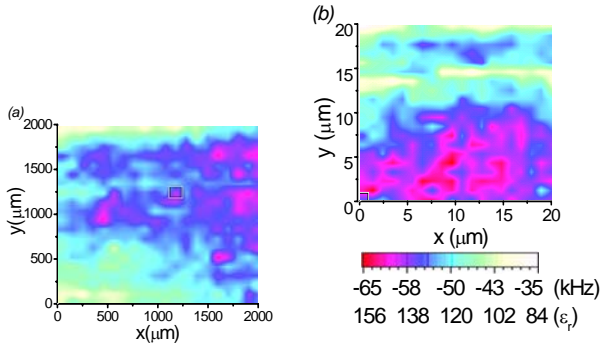

Fig 4. (a) Linear dielectric constant image of BTO-CFO thin film, 2000 μm x 2000 μm scanning area with 5 μm step size, and (b) 20 μm x 20 μm area scanning image with 0.5 μm step size, both taken at 3.8 GHz and room temperature. Color bar shows measured frequency shift and relative dielectric permittivity values.

With this localized dielectric measurement technique, we also have the possibility of measuring the magneto-electric effect locally. Because applying a magnetic field would induce a magnetostriction of the CFO phase, this strain is transferred to BTO, which is equivalent to the stress effect on BTO. Our microscope has a sensitivity of 2 kHz in frequency shift, which corresponds to $\Delta\varepsilon_r \sim 4$ in the 200-1000 relative dielectric constant range. Based on piezoelectric and magnetostriction coefficients ($e_{33}=18.6 C/m^2, q_{33}=699.7 N/Am$) [12], we estimate that a magnetic field of 3000 Oersted is needed to get a measurable dielectric change in the BTO-CFO film. The advantage for the proposed magneto-electric measurement is that we can localize it to the μm scale. This provides the capability to investigate the relation between magneto-electric properties and local composition or nano-structure.

## 3. Temperature Dependence of Linear Dielectric properties

The frequency shifts across the edge from MgO to BTO-CFO or BTO were recorded as a function of sample temperature from 30°C~200°C. Fig 5 show the temperature dependence of the frequency shift, which can be converted to temperature dependence of dielectric permittivity (we used the linear fit of ε~Δf curve shown in Fig.1 b). Due to the lattice transition of the perovskite structure of the BTO crystal, the relative dielectric permittivity should have a peak at the Curie temperature [13]. Firstly, we used a BTO bulk single crystal to verify this trend and to calibrate the thermometer from the transition temperature. Fig. 5(a) inset shows the dielectric change of the BTO crystal with temperature. There is a peak at 120°C, which is consistent with the Curie temperature from the literature [14]. The room temperature frequency shift corresponds to a permittivity of $\varepsilon_r$ ~ 800 (using the simulation curve for bulk samples in Fig. 1(a)), which is on the same order as the results for the thin film BTO sample.

BTO and BTO-CFO thin films on MgO also show a peak in $\varepsilon_r(T)$ which corresponds to the lattice transition of the BTO phase. However the broadening and flattening of the peak is typical of diffuse transitions, which can be attributed to structural inhomogeneity resulting from reduced grain size [13] and the strain gradient induced by the epitaxial growth of the thin films [15]. The broadened peak can be fit using a generalized Curie-Weiss equation [16]:

$$\frac{1}{\varepsilon} = \frac{1}{\varepsilon_{max}} + \frac{(T-T_{max})^2}{2\varepsilon_{max}\delta^2},$$

where the transition temperature is defined by $T_{max}$, which corresponds to the maximum dielectric constant $\varepsilon_{max}$, and δ is the diffuseness coefficient. Large values of the coefficient δ indicate a strongly diffuse transition. Here we found δ= 94 °C and transition temperatures $T_{max}$ = 154°C for the BTO thin film. The transition temperature shift is due to the stress from the substrate [11, 15]. The BTO-CFO doesn't show the same enhanced transition temperature as the BTO film. From Fig. 5(b), the transition temperature of BTO-CFO is about 124°C which is similar to bulk BTO. The CFO thin film on MgO does not show an obvious temperature dependence in $\varepsilon_r(T)$ in this range (not shown).

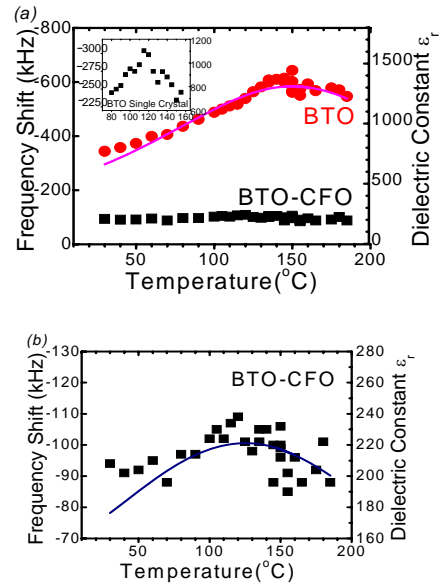

Fig 5. (a) Temperature dependence of dielectric constant for BTO thin film (red circles), BTO-CFO thin film (black squares), and bulk BTO single crystal (inset); (b) enlargement of BTO-CFO curve in (a). Also shown as solid lines on BTO and BTO-CFO are fits to the generalized Curie-Weiss equation discussed in the text. All data taken at 3.8 GHz. Note that the frequency shift is plotted with a sign opposite to that in Figs. 1 and 3.

### 4. Non-linear Dielectric Properties

The dielectric tunability measurements were done at a single point on similar thin films but with an SRO counterelectrode present (Fig 2.b). A 1 V dc bias corresponds to a local electric field on the scale of 60 kV/cm in the dielectric film. Fig. 6 (inset) is a schematic of a typical frequency shift curve as a function of the bias voltage on a BTO-CFO film. The results show that BTO-CFO has a non-linear dielectric property similar to thin film BTO. Applying a bias voltage, the dielectric permittivity is reduced, so that around 0 bias voltage the relative permittivity has a maximum value.[10] Further measurements on CFO/SRO as a function of bias voltage shows only frequency shift changes at the noise level of the experiment. Hence we conclude that the non-linearity in BTO-CFO/SRO comes from the BTO phase and not the SRO layer. The peak in the tunability curve corresponds to the reversal of polarization (coercive field), and the amplitude corresponds to the amount of switchable polarization [17]. With decreasing temperature below $T_c$, the tunability curve will exhibit a more pronounced hysteresis in which the upgoing and downgoing frequency shift curves cross. Above $T_c$, the BTO is in the paraelectric state and the tunability curves exhibit weakly hysteretic behavior. [18] In our experiment, this transition occurs around 130°C for both the BTO-CFO and BTO films (see Fig. 6).

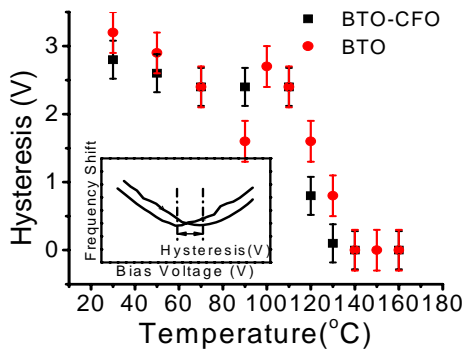

Figure 6. Hysteresis measured at 3.8 GHz. Both BTO (red circles) and BTO-CFO (black squares) show a critical temperature of ~130°C at which the hysteresis reduces to zero. The inset schematic shows a typical frequency shift change as a function of bias voltage for the configuration shown in Fig. 2(b). The hysteresis is defined as the voltage difference between two extrema measured in upward and downward voltage sweeps.

Table I summarizes the samples used in the experiment and the main results for the dielectric properties. For pure BTO, the transition temperature determined from linear dielectric measurements is different from that determined by non-linearity measurements. We believe the shift of the BTO Curie temperature is caused by the substrate clamping effect. Introducing an SRO counter-electrode layer would relax the stress on the film thus changing the Curie temperature of the film towards to that of bulk BTO. For BTO-CFO films the transition temperature determined from the dielectric measurement is about the same as that from the non-linear measurements, showing that the properties of BTO-CFO are less dependent on the underlying layer. The transition temperature shift of BTO-CFO is less pronounced, and if it exists at all, would be mainly caused by the CFO phase, either through stress caused by lattice mismatch or Co and Fe diffusing into the BTO phase leading to a metastable solid solution.[5]

### Summary

In summary, we measured the linear and non-linear dielectric properties of BTO-CFO nano-composites. The dielectric properties of BTO-CFO are similar to those of BTO including the temperature dependence and bias voltage dependence. The measurement has a quantitative spatial resolution of approximately 5 μm. The dielectric change as well as the transition temperature change in the BTO component of BTO-CFO is mainly due to the interaction of BTO with CFO. The possibility of localized magneto-electric measurement with this technique is also discussed.

**Acknowledgements**:
We gratefully acknowledge assistance from Alexander Tselev and Atif Imtiaz. This work was supported by the NSF MRSEC under contracts NSF DMR-00-80008 and NSF DMR-05-20471, as well as NSF ECS-0322844 and NSF DMR-0302596.

**References**:
[1] Wood, V. E. & Austin, A. E. Possible applications for magnetoelectric materials. *Int. J. Magn.* **5**, 303–315 (1974) .
[2] G. A. Smolenskii and I. E. Chupis, Ferroelectromagnets. Sov. Phys. Usp. **25**, 475 (1982).
[3] B. I. Al'shin and D.N. Astrov, Magnetoelectric effect in Titanium Oxide $Ti_2O_3$. Soviet Physics JETP 17, 809 (1963).
[4] J. G. Wan, J.-M. Liu, H. L. W. Chand, C. L. Choy, G. H. Wang, and C. W. Nan, Giant magnetoelectric effect of a hybrid of magnetostrictive and piezoelectric composites. J.

Appl. Phys. **93**, 9916 (2003).

[5] K.-S. Chang, M. A. Aronova, C.-L. Lin, M. Murakami, M.-H. Yu, J. Hattrick-Simpers, O. O. Famodu, S. Y. Lee, R. Ramesh, M. Wuttig, I. Takeuchi, C. Gao, and L. A. Bendersky, Exploration of Artificial Multiferroic Thin Film Heterostructures Using Composition Spreads, Appl. Phys. Lett. **84**, 3091 (2004).

[6] H. Zheng, J. Wang, S. E. Lofland, Z. Ma, L. Mohaddes-Ardabili, T. Zhao, L. Salamanca-Riba, S. R. Shinde, S. B. Ogale, F. Bai, D. Viehland, Y. Jia, D. G. Schlom, M. Wuttig, A. Roytburd, R. Ramesh, Multiferroic $BaTiO_3$-$CoFe_2O_4$ nanostructures, Science **303**, 661 (2004).

[7] D. E. Steinhauer, C.P. Vlahacos, F.C. Wellswood, and S. M. Anlage, Quantitative imaging of dielectric permittivity and tunability with a near field scanning microwave microscope, Rev. Sci. Instrum. **71**, 2751 (2000).

[8] H. M. Altshuler, in Handbook of Microwave Measurements II, edited by M. Sucher and J. Fox (Polytechnic Inst. Of Brooklyn, Brooklyn, NY 1962) p. 225.

[9] Atif Imtiaz, Marc Pollak, Steven M. Anlage, John D. Barry and John Melngailis, Near-field microwave microscopy on nanometer length scales**,** J. Appl. Phys. **97**, 044302 (2005); Atif Imtiaz, and Steven M. Anlage, Effect of tip-geometry on contrast and spatial-resolution of the near-field microwave microscope**,** J. Appl. Phys., **100,** 044304 ( 2006).

[10] D.E. Steinhauer, C.P. Vlahacos, F.C. Wellstood, S.M. Anlage, C. Canedy, R. Ramesh, A. Stanishevsky and J. Melngailis, Imaging of microwave permittivity, tunability, and damage recovery in $(Ba,Sr)TiO_3$ thin films, Appl. Phys. Lett. **75**, 3180 (1999)

[11] Steven M. Anlage, Vladimir V. Talanov, and Andrew R. Schwartz, Principles of Near-Field Microwave Microscopy, section 4.7, *Scanning Probe Microscopy: Electrical and Electromechanical Phenomena at the Nanoscale*, edited by S. Kalinin and A. Gruverman (Springer, New York, 2006), pp. 207-245.

[12] Ce-Wen Nan, Magneto-electric effect in composites of piezoelectric and piezomagnetic phases, Phys. Rev. B **50**, 6082 (1994).

[13] B. H. Hoerman, G. M. Ford, L.D. Kaufmann, and B.W. Wessels, Dielectric properties of epitaxial $BaTiO_3$ thin films, Appl. Phys. Lett. **73**, 16 (1998).

[14] Sakayori K, Matsui Y, Abe H, Japanese Journal of Applied Physics Part 1 **34**(9B), 5443-5445 (1995).

[15] Sinnamon, L. J., Bowman, R. M. & Gregg, J. M. Thickness-induced stabilization of ferroelectricity in $SrRuO_3$/$Ba_{0.5}Sr_{0.5}TiO_3$/Au thin film capacitors. Appl. Phys. Lett*,* **81**, 889-891 (2002).

[16] C. B. Parker, J.-P. Maria, and A. I. Kingon, Temperature and thickness dependent permittivity of $(Ba,Sr)TiO_3$ thin films, Appl. Phys. Lett. **81**, 340 (2002).

[17] S. L. Miller, R. D. Nasby, J. R. Schwank, M. S. Rodgers, and P. V. Dressendorfer, Device modeling of ferroelectric capacitors, J. Appl. Phys. **68**, 6463 (1990).

[18] J. Miao, H. Yang, W. Hao, J. Yuan, B. Xu, X. Q. Qiu, L. X. Cao and B. R. Zhao, Temperature dependence of the ferroelectric and dielectric properties of the BST/LSMO heterostructure, J. Phys. D: Appl. Phys. **38**, 5 (2005).